\documentclass[aps,pra,showpacs,twocolumn,showkeys,amssymb,amsmath,superscriptaddress,reprint]{revtex4-1}

\usepackage{lipsum}
\usepackage{amsmath}
\usepackage{graphicx}
\usepackage{color}
\usepackage{tikz}
\usepackage{verbatim}

\usepackage{amssymb}
\usepackage{bm}

\newcommand{\rom}[1]{\textup{\uppercase\expandafter{\romannumeral#1}}}

\begin{document}

\title{Characteristic length scales from entanglement dynamics in electric-field-driven tight-binding chains}

\author{Devendra Singh Bhakuni}
\email{devendra123@iiserb.ac.in}
\author{Auditya Sharma}
\email{auditya@iiserb.ac.in}
\affiliation{Department of Physics, Indian Institute of Science Education and Research, Bhopal, India}


\begin{abstract}
We study entanglement dynamics in the nearest-neighbour fermionic
chain that is subjected to both DC and AC electric fields. The dynamics 
gives the well known Bloch oscillations in the DC field case provided 
that the system size is larger than the Bloch length whereas in the 
AC field case the entropy is bounded and oscillates with the driving frequency 
at the points of dynamic localization, and has a logarithmic growth at other points. 
A combined AC + DC field yields super Bloch oscillations for the system size
larger than the super Bloch length which puts a constraint on the
device size in a typical non-equilibrium set-up to observe super Bloch
oscillations where the device is connected to the leads. Entanglement entropy provides
useful signatures for all of these phenomena, and an alternate way to capture the various length scales involved.
\end{abstract}

\maketitle

\section{Introduction}
A charged particle under the influence of a static electric field
would be expected to undergo accelerated motion towards infinity.
However, in the presence of a periodic lattice potential and electric
field the motion of a quantum particle is oscillatory. These
oscillations are known as Bloch oscillations~\cite{zener1934theory,wannier1960wave,krieger1986time,bouchard1995bloch}.
The time period of these oscillations is inversely proportional to the
field strength. Although theoretically well-understood, Bloch oscillations 
are hard to control experimentally in normal crystals
where lattice imperfections often wash out this effect altogether~\cite{hofmann2015solid}. However, the advent 
of semiconductor super-lattices, optical lattices, and temperature-tuned wave guides has made
it possible to realize Bloch oscillations experimentally~\cite{mendez1993wannier,waschke1994experimental,
  dekorsy1994terahertz,dahan1996bloch,morsch2001bloch}.

The study of Bloch oscillations in a tight-binding framework gives the
well known Wannier-Stark ladder as the energies and Wannier-Stark
states as the eigenstates of the
Hamiltonian~\cite{hartmann2004dynamics}. These states are extended
over a length called \emph{Bloch length}. Hence, to observe Bloch
oscillations, the system size must be greater than this
length~\cite{popescu2017emergence}. Also, the equispaced energy
spectrum results in recurrence of any initial state. Thus an
initially localized state will again be localized after a Bloch period.

By making the electric field time dependent: $\mathcal{F} (t) = A \cos
\omega t$, with time period $T=2\pi/\omega$, one would expect that the
same periodicity can be seen in the dynamics. However contrary to this expectation,
only at certain special ratios of the amplitude and the frequency of the
drive the periodicity can be seen. This phenomenon is called
\emph{dynamic localization}~\cite{dunlap1986dynamic,arlinghaus2011dynamic,
  eckardt2009exploring} and the special ratios are the roots of Bessel
function of order zero.

A more general form of the electric field, which includes both AC and
DC field gives other interesting effects such as \emph{coherent
  destruction of Wannier-Stark
  localization}~\cite{holthaus1995random,holthaus1995ac}, which occurs
when the DC field is resonantly tuned with the AC drive, and 
\emph{super Bloch
  oscillations}~\cite{kudo2011theoretical,kolovsky2010dynamics,
  caetano2011wave} which occurs for slight detuning from the
resonant drive. The time period of these oscillations is very large in
comparison to Bloch oscillations. Similar to the static field case, a
\emph{super Bloch length}~\cite{hu2013dynamics,haller2010inducing} is
associated with the super Bloch oscillations and in order to observe
these oscillations the system size must be greater than this length.

Entanglement~\cite{laflorencie2016quantum,amico2008entanglement,peschel2009reduced}
quantifies the quantum correlations in a state between two parts of a
system. In recent times, it has emerged as a unique tool to capture a
vast variety of phenomena ranging from quantum phase
transitions~\cite{osborne2002entanglement,osterloh2002scaling,vidal2003entanglement},
localization/de-localization transitions~\cite{roy2018entanglement,roosz2014nonequilibrium,PhysRevLett.115.046603,PhysRevLett.110.260601,PhysRevLett.109.017202}, trivial to topological
transition~\cite{nehra2018many,sirker2014boundary,cho2017quantum} etc. in closed systems to the
correlations between a quantum dot and the baths in open quantum
systems~\cite{sharma2015landauer,sable2018landauer}.

The entanglement in a Wannier-Stark ladder~\cite{eisler2009entanglement}
and various types of quench dynamics has been studied before, both
numerically~\cite{eisler2008entanglement,eisler2007evolution}, and
using CFT calculations~\cite{calabrese2007entanglement}. However, the
connection to various phenomena associated with electric field was
not made explicitly. From the time evolution of an initially half
filled state where all the particles are filled in the left half of
the chain, we find the appearance of Bloch oscillations and super
Bloch oscillations in the entropy dynamics when the system size is
larger than the respective length scales: Bloch length and super Bloch
length. Hence, a device length greater than the super Bloch length is
required if one has to observe super Bloch oscillations in a
non-equilibrium set-up where the device is connected to
metallic leads fixed at different chemical potentials and temperatures.
Also, oscillatory behaviour of the entropy is observed at the
dynamically localized point whereas a logarithmic growth in entropy is
seen for the situations where a resonant drive destroys the Wannier-Stark localization.

To the best of our knowledge, the extraction of the various length scales that arise 
in the dynamics of a tight-binding chain subjected to electric field, from a study of entanglement is being reported for
the first time in this paper. Furthermore, well-known associated phenomena like Bloch oscillations, dynamic localization, coherent destruction of Wannier-Stark localization
and super-Bloch oscillations are viewed from an entanglement perspective.
The organization of this paper is as follows. In the next
section, we describe the model Hamiltonian with a general time-dependent field. 
The subsections present the different forms of the
field and the associated phenomena. The following section provides the
quench protocol and the entanglement dynamics with numerical
results. The summary and conclusions are given in the last section.

\section{Model Hamiltonian} \label{Model hamiltonian}
The Hamiltonian for a $1\text{D}$ tight binding model, for a finite
system of size $2L$ sites with electric field is
\begin{equation}
H=-J\sum_{n=-L+1}^{L-1}c_{n}^{\dagger}c_{n+1}+c_{n+1}^{\dagger}c_{n}+a\mathcal{F}(t)\sum_{n=-L+1}^{L}(n-1/2) c_{n}^{\dagger}c_{n},\ \quad
\end{equation}
where $\mathcal{F}(t)$ is the electric field and $a$ is the lattice parameter. The numerical work is done in units where
$\hbar = 1, c=1, e=1, J= 1,a=1$ (unless otherwise stated). For a constant electric field
$\mathcal{F}(t)=F$, the dynamics gives the well known Bloch
oscillations, while a sinusoidal driving $\mathcal{F}(t)=A\cos\omega
t$ can give rise to dynamic localization when $A$ and $\omega$ are
tuned appropriately. In the presence of a combined AC+DC field
$\mathcal{F}(t)=F+A\cos\omega t$, the phenomenon of coherent
destruction of Wannier-Stark localization is seen at resonance whereas
a slight detuning from the resonant condition yields super Bloch
oscillations.

\subsection{Bloch Oscillations}
For $\mathcal{F}(t)=F$, the exact eigenstates in the $n\rightarrow \infty$ limit are the  Wannier-Stark states:
\begin{equation}
|\Psi_{m}\rangle = \sum_{m} \mathcal{J}_{n-m}\left(2J/aF\right)|n\rangle.
\end{equation}
The single particle energies form a ladder with equal spacing
$E_{m}=maF$, where $m=0,\pm 1,\pm 2, .... $. However, for finite system
sizes one observes non-linear behaviour at the ends of the spectrum - these end effects are diminished on increasing system size. 

\begin{figure}[t]
\includegraphics[scale=1.1]{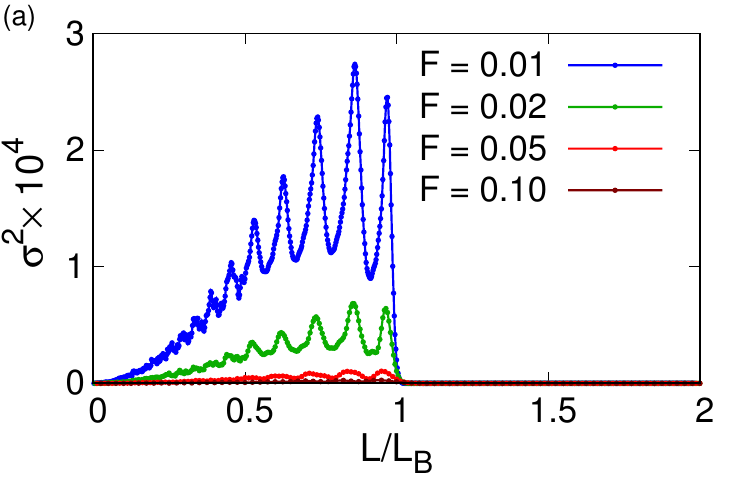}
\includegraphics[scale=1.1]{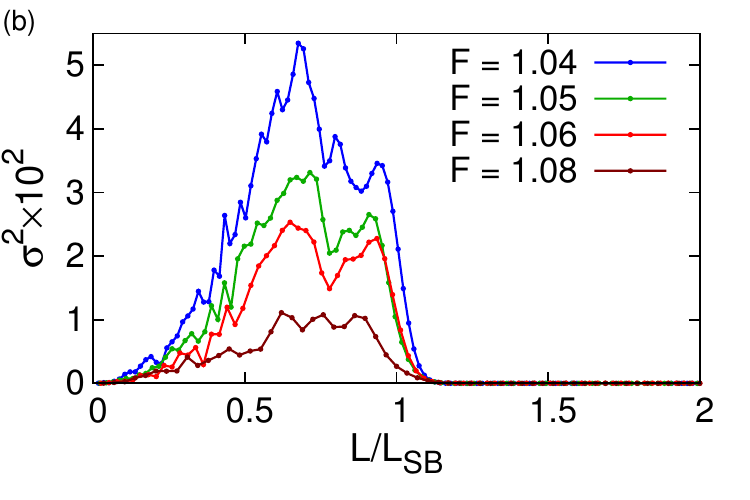}
\caption{Mean square single particle width of the wave-packet at $t=5T_B$ (top) and $t=2T_\text{SB}$ (bottom) as a function of system size rescaled to Bloch length and super Bloch length respectively. The transition shows the requirement of a device size greater than Bloch length and super Bloch length to observe the oscillations.(Parameters are: $A=2.0$ and $\omega=1.0$.)}
\label{widthBL}
\end{figure}

A natural length scale of the problem is the Bloch length~\cite{hartmann2004dynamics}
\begin{equation}
L_{B}=\frac{4J}{F}.
\end{equation}
One can observe Bloch oscillations with the frequency $\omega_B = aF/\hbar$
in the dynamics if the system size is greater than the Bloch
length. This can be easily seen by studying the time evolution of the mean square single particle width
\begin{equation}
\sigma^2(t) = \langle n^2(t)\rangle - \langle n(t)\rangle ^2,
\end{equation}
when the initial state is a single particle localized at the centre of
the chain. Time evolution for a time equal to an integer multiple of $T_B$ must return the system back to
where it started, if Bloch oscillations are present. Fig.~\ref{widthBL}(a) shows the mean square width of the wave-packet
calculated for different field strengths at a time which is an
integer multiple of $T_B$ (we show data for the specific case of
$5T_B$), as a function of the system size rescaled by the Bloch
length. Since the initial state is a localized state with
$\sigma^2=0$ and the same state will reappear after a Bloch period
(see appendix), the mean square width at $t=nT_B$ should also go to zero. However,
this happens only when the system size is larger than the Bloch
length.
\subsection{Dynamic Localization}
For a time dependent electric field, the solution of the time
dependent Schr\"{o}dinger equation for the Hamiltonian $H$ is given
by the Houston states or accelerated Bloch states~\cite{houston1940acceleration,arlinghaus2011dynamic} as
\begin{equation}\label{houston}
|\psi_{k}(t)\rangle = \exp\left(-\frac{i}{\hbar}\int_{0}^{t}E(q_{k}(\tau))d\tau\right)\sum_{n}|n\rangle \exp(inq_{k}(t)a),
\end{equation}
where the quasi-momentum's time dependence can be obtained from the equation of motion as
\begin{equation}
q_{k}(t)=k+\frac{1}{\hbar}\int_{0}^{t}d\tau \mathcal{F}(\tau).
\end{equation}
The dispersion $E(q_{k})\equiv -2J\cos(q_{k}a)$ is that of the nearest-neigbour tight-binding model, but with implicit time-dependence from the quasi-momentum.
For a sinusoidal driving $\mathcal{F}(t)=A\cos(\omega t)$, the quasi-momentum is
\begin{equation}
q_{k}(t)=k+\frac{A}{\hbar\omega}\sin(\omega t).
\end{equation}
Although the quasi momentum $q_{k}(t)$ is periodic in time with period
$T=2\pi/\omega$, the Houston states however are \emph{not}
$T$-periodic due to an extra contribution coming from the
integral appearing in the exponential. The one cycle average of that
contribution is
\begin{eqnarray}
\epsilon(k)=\frac{1}{T} \int_{0}^{T} d\tau E(q_{k}(t))\quad\qquad\qquad\qquad\quad\nonumber \\
= \frac{-2J}{T} \int_{0}^{T} d\tau \cos\left(ka+\frac{Aa}{\hbar\omega}\sin(\omega t)\right)  \nonumber \\
=-2J\mathcal{J}_{0}\left(\frac{Aa}{\hbar\omega}\right)\cos(ka)\nonumber \quad\qquad\qquad\ \ \ \\
=-2J_{\text{eff}}\cos(ka)\quad\qquad\qquad\qquad\ \ \ \ \ \ \ .
\end{eqnarray}
The effect of the drive is thus the well-known re-normalization of the
hopping parameter. The Houston states can now be decomposed in
Floquet notation with the quasi-energies $\epsilon(k)$ as
\begin{equation}
|\psi_{k}(t)\rangle = |u_{k}(t)\rangle \exp\left(-\frac{i}{\hbar}\epsilon(k)t\right),
\end{equation}
where $|u_{k}(t)\rangle$ \emph{is} $T$-periodic function which is obtained by removing the extra contribution term of the exponential.

Hence, the time evolution of any initial state is given by a periodic
function $u_{k}(t)$ and the phase factors $\exp\left(-i\epsilon(k)t/\hbar\right)$. 
However, the phase can be tuned to unity by choosing $(\frac{Aa}{\hbar \omega})$ to be a zero of the
Bessel function of order zero thereby collapsing the band. In such
a scenario, the initial state will reappear after time $T$ and a
localized wave-packet remains localized. This phenomenon is known as
dynamic localization.

\subsection{Coherent Destruction of Wannier-Stark Localization and Super Bloch Oscillations }
Although a static electric field destroys band formation and leads
to Wannier-Stark localization and an AC driving gives dynamic
localization, a combination of these two can give rise to the counter-intuitive phenomenon of 
coherent destruction of Wannier-Stark localization. At resonant tuning, the effective model is the nearest-neigbor tight-binding model, and
an initially localized wavepacket delocalizes according to the usual mechanism. When a slight 
detuning from resonance is imposed, super Bloch oscillations whose frequency is determined by the detuning are seen.

The formalism is the same as given in the previous section, however the net force is now $\mathcal{F}(t)=F + A\cos(\omega t)$. 
The quasi momentum given by
\begin{equation}
q_{k}(t)=k+Ft+\frac{A}{\hbar\omega}\sin(\omega t)
\end{equation}
is \textit{not} $T$-periodic except at resonance: 
\begin{equation}
Fa=n\omega,
\end{equation}
where $n$ is an integer. 
\begin{eqnarray}
q_{k}(t+T)=k+\frac{2n\pi}{aT} t+\frac{2n\pi}{a}+\frac{A}{\hbar\omega}\sin(\omega t)\nonumber \\
=k+\frac{2n\pi}{aT} t+\frac{A}{\hbar\omega}\sin(\omega t)\nonumber \qquad\quad\\
=q_{k}(t),\qquad\quad\qquad\quad\qquad\quad \quad \ \ 
\end{eqnarray}
where, the periodicity of the lattice vector is used in the last step.
\begin{figure}[t]
\includegraphics[scale=1.1]{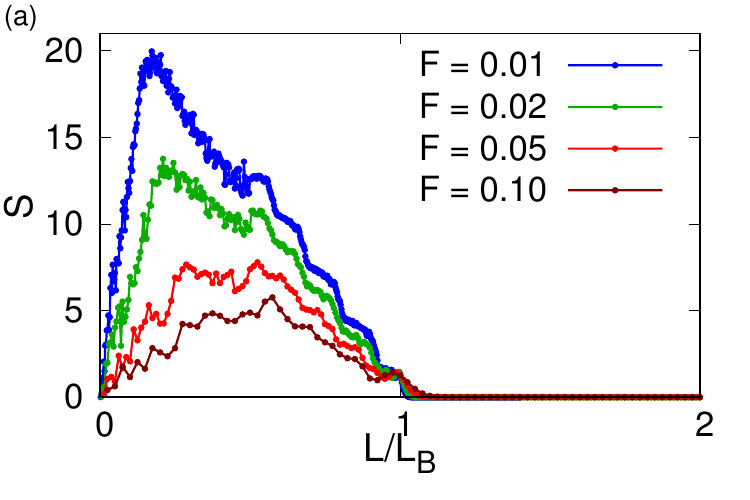}
\includegraphics[scale=1.1]{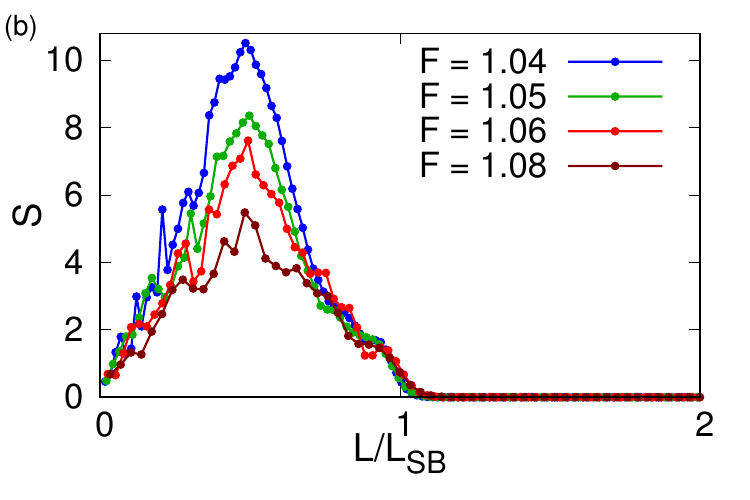}
\caption{Entanglement entropy as a function of $L/L_B$ and $L/L_{SB}$  for a DC field (top) and combined AC+DC field with slight detuning (bottom) at $t=5T_B$ and $t=3T_{SB}$ ($A=2.0$ and $\omega=1.0$) . The initial state is a half-filled state where all the particles are filled on the left half of the chain. The entropy becomes zero when the system size is larger than the Bloch length and super Bloch length, respectively. }
\label{entBL}
\end{figure} 
The quasi-energies can be calculated as
\begin{eqnarray}
\epsilon(k)=\frac{1}{T} \int_{0}^{T} d\tau E(q_{k}(t))\quad\qquad\qquad\qquad\quad\qquad\quad\nonumber \\
= \frac{-2J}{T} \int_{0}^{T} d\tau \cos\left(ka+n\omega t + \frac{Aa}{\hbar\omega}\sin(\omega t)\right)\   \nonumber \\
=(-1)^{n}\mathcal{J}_{n}\left(\frac{Aa}{\hbar\omega}\right)(-2J\cos(ka))\nonumber \quad\qquad\qquad \ \ \\
=-2J_{\text{eff}}\cos(ka).\quad \quad \quad\qquad\qquad\qquad\qquad\quad \ 
\end{eqnarray}
It can be seen that AC driving leads to the formation of
bands with the hopping parameter getting renormalized. Hence under such a
resonant driving the particle can delocalize even though the static
electric field is present. However, once again at the zeros of the Bessel function of
order $n$, there is dynamic localization.
\begin{figure*}[t]
\includegraphics[scale=0.75]{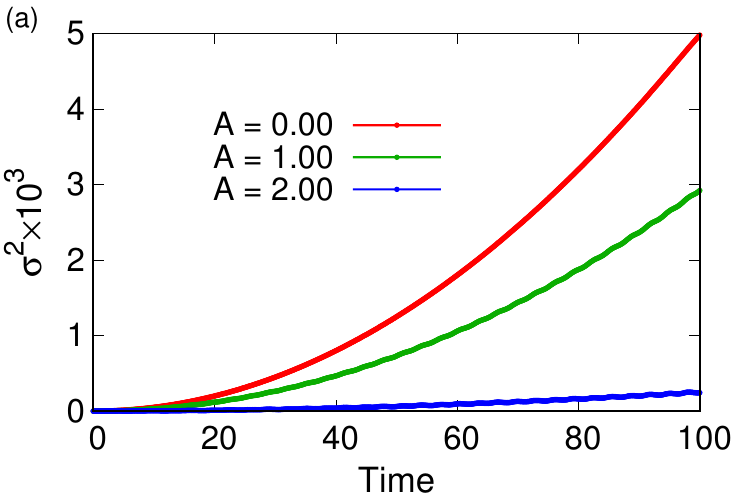}
\includegraphics[scale=0.75]{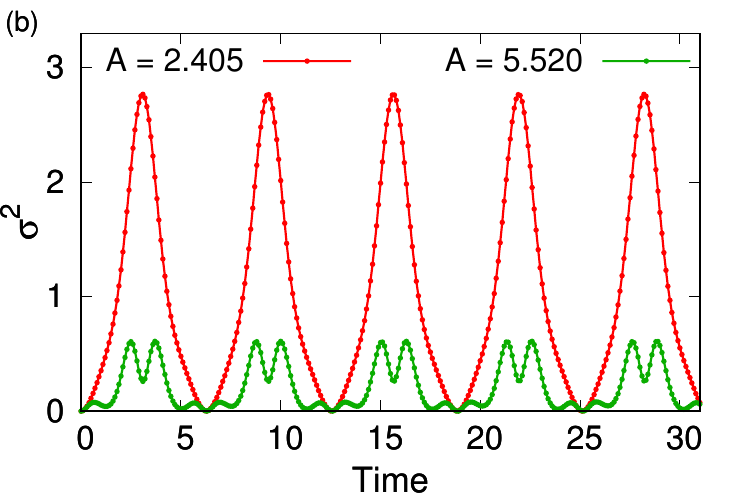}
\includegraphics[scale=0.75]{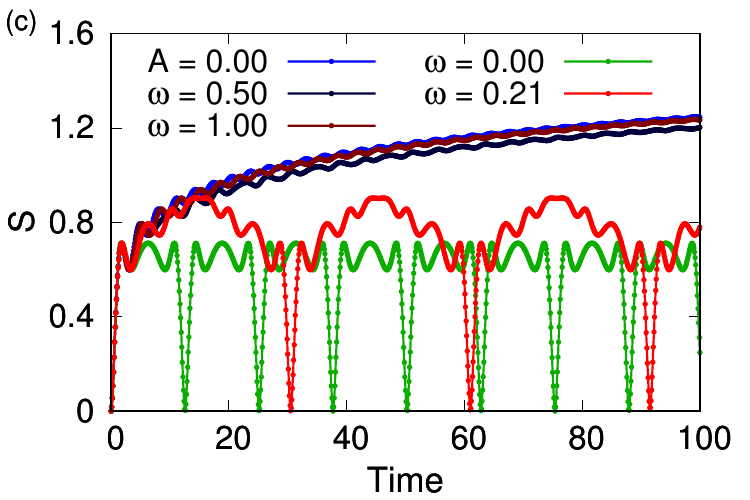}
\caption{Mean square width of an initially localized wave packet for
  an AC driving. Mean square width is unbounded for arbitrary
  $A/\omega$ (left), whereas it is bounded and periodic when $A/\omega$ is a
  root of the Bessel function of zeroth order (centre)(parameters are
  $L=100$ and $\omega = 1.0$). Entanglement entropy (right) for half
  filled localized state. For $\omega=0.0$, the entropy shows Bloch
  oscillations, whereas for other values the entropy is
  bounded/unbounded depending on the ratio of $A/\omega$. (Parameters
  are $L=100$, $J=0.5$, and $A=0.5$ for all curves except the blue one.)}
\label{entvar}
\end{figure*} 

A slight detuning from the resonant drive gives rise to super Bloch oscillations. The off-drive condition can be written as 
\begin{equation}
Fa=(n+\delta)\omega.
\end{equation}
Under this condition an initially localized wave-packet starts
oscillating with the time period $T_{SB}=\frac{2\pi}{\delta\omega}$~\cite{kudo2011theoretical,PhysRevB.86.075143}. These oscillations
are known as super Bloch oscillations.  Similar to the Bloch length, a
super Bloch length is associated with super Bloch oscillations
\begin{equation}
L_\text{SB} = \frac{J \mathcal{J}_{n}(\frac{Aa}{\hbar\omega})}{\delta\omega}.
\end{equation}
The super Bloch oscillations can be seen if the system size is greater than this length. The same is shown in Fig.~\ref{widthBL}(b) where the wave-packet mean square width is calculated as a function of $L/L_{SB}$.

To summarize this section, we have made a review of the phenomena of Bloch oscillations, dynamic localization, coherent destruction of Wannier-Stark localization, and super Bloch oscillations. We have shown how
the length scales $L_{B}$ and $L_{SB}$ can be extracted from a study of wave-packet width. This section sets the scene for how these phenomena may be viewed from an entanglement perspective 
in the next section.

\section{Entanglement entropy and quench dynamics}
Quantum entanglement quantifies the lack of information of any
subsystem despite full knowledge of the overall system, and is also a measure of how one part of
the system is correlated with another part. Although
various measures of entanglement are available in the literature~\cite{eisert2010colloquium,vidal2002computable,wootters1998entanglement,sharma2015landauer}, the one
that is most widely used as an entanglement measure is the von Neumann
entropy.
 
Let $\rho$ be the density matrix which describes the full state of the system. The von Neumann entropy
of the any subsystem $A$ is defined as
\begin{equation}
S_{A}=-\text{Tr}(\rho_{A}\text{log}\rho_{A}),
\end{equation}
where $\rho_{A}$ is the reduced density matrix of subsystem $A$,
i.e. $\rho_A = \text{Tr}_B(\rho)$ with $\text{Tr}_B$ denoting the
partial trace with respect to subsystem $B$. When the overall state $\rho$ is pure, the von Neumann entropy $S_{A}$ is also
the entanglement entropy between $A$ and $B$.

In a non-interacting quadratic fermionic system, the von Neumann
entropy can be directly computed from the two point correlation
matrix~\cite{peschel2004reduced} of the subsystem $A$: $C_{mn}=\langle
c_{m}^{\dagger}c_{n}\rangle$. The von Neumann entropy is given in terms
of the eigenvalues $n_{\alpha}$ of the subsystem correlation matrix as
\begin{equation}
\label{eq:entropy}
S=\sum_{\alpha}\left[-\left(1- n_{\alpha}\right)\text{ln}\left(1- n_{\alpha}\right)-n_{\alpha}\text{ln}\ n_{\alpha}\right].
\end{equation}
A generalization of the above result facilitates the study of the
dynamics of entanglement.  The system is initially prepared in the
ground state of an unperturbed Hamiltonian and then suddenly a
suitable quenching Hamiltonian is switched on, and the resulting time-evolution
governed by the new Hamiltonian is tracked. The time dependent
correlation matrix is then directly constructed from the time
evolution of the state. Finally, from the time-dependent eigenvalues
of the subsystem correlation matrix, we have access to the dynamics of 
entanglement entropy.

The time evolution of the initial state is given by 
\begin{equation}
|\psi(t)\rangle=e^{-iHt/\hbar}|\psi(0)\rangle
\end{equation}
The time dependent correlation matrix can be written as 
\begin{eqnarray}
C_{mn}(t)=\langle \psi(t)|c^{\dagger}_{m}c_{n}|\psi(t)\rangle=\langle \psi_0|c^{\dagger}_{m}(t)c_{n}(t)|\psi_0\rangle,\quad
\end{eqnarray}
where we have switched to the Heisenberg picture. Using the time
evolution of fermionic operators $c^{\dagger}_{j}$ and $c_{j}$, we can
simplify the expression for correlation matrix as (see appendix)
\begin{equation}
C(t)=U^{\dagger}(t)C(0)U(t),
\end{equation} 
where $U_{jk}(t)=\sum_{n}D^*_{jn}\exp(-i\epsilon_{n}t/\hbar)D_{nk}$ and the matrix $D$ diagonalizes the new Hamiltonian.

In a quenching protocol, where the final Hamiltonian includes a static electric field, the form of the unitary matrix $U$ can be written as~\cite{hartmann2004dynamics}
\begin{equation}
U_{n n^\prime}\left(t\right)=\mathcal{J}_{n-n^{\prime}}\left(\frac{4J}{\hbar\omega_B}\sin{\frac{\omega_{B}t}{2}}\right)e^{i\left(n-n^{\prime}\right)\left(\pi-\omega_{B}t\right)/2-in^\prime\omega_{B}t},
\end{equation}
Since, $U(t)$ is periodic in time with the Bloch period, the correlation matrix also follows the same periodicity.
Now, the correlation matrix can be written as
\begin{widetext}
\[
C_{mn}=\sum_{qq^{\prime}}\mathcal{J}_{q-m}\left(\frac{4J}{\hbar\omega_B}\sin{\frac{\omega_{B}t}{2}}\right)e^{-i\left(q-m \right)\left(\pi-\omega_{B}t\right)/2+i m\omega_{B}t} \ C_{qq^{\prime}}(0) \ 
\mathcal{J}_{q^\prime-n}\left(\frac{4J}{\hbar\omega_B}\sin{\frac{\omega_{B}t}{2}}\right)e^{-i\left(q^\prime-n \right)\left(\pi-\omega_{B}t\right)/2-i n\omega_{B}t}, \qquad\qquad\qquad
\]
\end{widetext}
which in conjunction with Eqn.~\ref{eq:entropy} yields the
time-dependent entanglement entropy.  For a time dependent
Hamiltonian, we discretize the time interval into tiny regions where
the Hamiltonian does not change appreciably, and the same procedure as
above follows within the tiny intervals. In this case, the time evolution operator
consists of a series of unitary operators.

The entanglement entropy between the two subsystems for the
Wannier-Stark problem was studied
in~\cite{eisler2009entanglement}. The contribution to the entanglement
entropy comes from the interface width which is created by the
potential gradient. Also the dynamics after turning off the electric
field was studied and a logarithmic growth of the entropy was
observed. However, here we entirely focus on the dynamics of the
entanglement entropy after the electric field is turned \emph{on}. The entropy
as a function of system size rescaled by the Bloch length and super
Bloch length at $t=5T_B$ and $t=3T_\text{SB}$ respectively, are plotted
in Fig.~\ref{entBL}. The initial state can be thought of as the ground
state of a half filled nearest neighbour tight binding chain with a
large electric field, where all the particles are localized to the
left of the chain. A transition signifies the minimum system size to
observe Bloch oscillations and super Bloch oscillations. The
requirement of device length larger than the Bloch length has been given
in recent work~\cite{popescu2017emergence} where the device is connected to two
metallic leads in a non-equilibrium setting and a transition from DC
regime to Bloch oscillations regime is observed on varying the device
size. Our finding is that a similar result holds for the super Bloch
oscillations in the presence of a combined AC and DC field where the
device size has to be greater than the super Bloch length.

\begin{figure*}[t]
\includegraphics[scale=0.76]{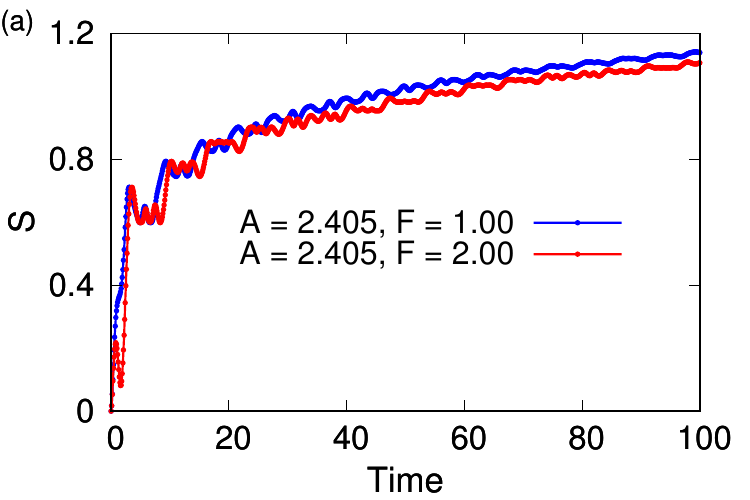}
\includegraphics[scale=0.76]{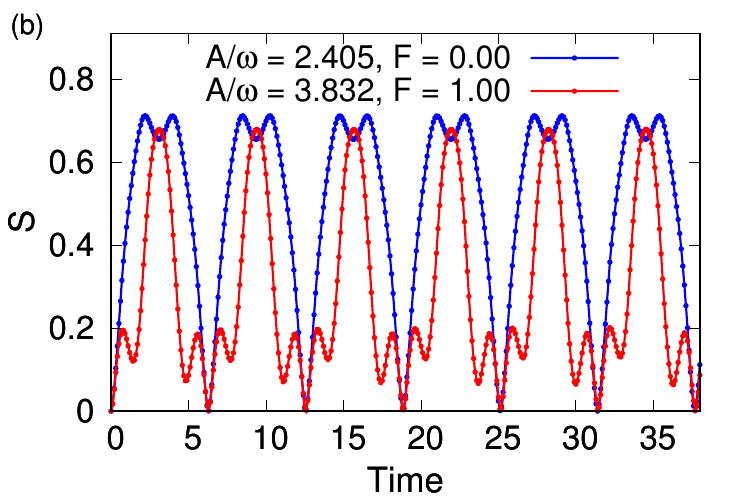}
\includegraphics[scale=0.76]{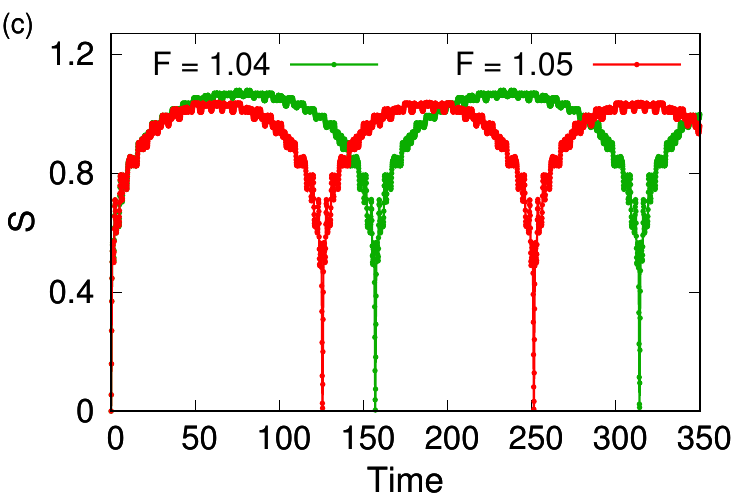}
\caption{Destruction of Wannier-Stark localization (left) for resonantly tuned driving $F=n\omega, n=1,2..$ and dynamic localization (centre) under the action of both static electric field and AC driving from the dynamics of entanglement entropy. Super Bloch oscillations captured by  entanglement entropy for a detuning from the resonant  driving(right). The parameters are $L=100$ for all plots $J=0.5$ for (a), (b) and $A=3.0$ for (c).  }
\label{entACDC}
\end{figure*}

For the case of AC driving, Fig.~\ref{entvar}(a,b) shows that the
mean square single-particle width of the wave-packet is unbounded in time at
arbitrary ratios of $A/\omega$. In contrast, it remains bounded and oscillatory
at the special ratios which are the roots of the Bessel function of
order zero. The time evolution of entropy is shown in
Fig.~\ref{entvar}(c) for various parameters. One can see the Bloch
oscillations in the dynamics of entropy for $\omega = 0.0$ and dynamic
localization for the ratio $A/\omega=2.405$, which is a root of Bessel
function of order zero. For all other cases the entropy is unbounded
in time, thus signifying de-localization. The coherent destruction
of Wannier-Stark localization is shown in Fig.~\ref{entACDC}(a) for a
combined AC and resonantly tuned AC field. The entropy keeps on
increasing for a resonantly tuned AC, even in the presence of the DC field. However
again the periodicity of the drive can be seen in the entropy at the
special ratios which are now the roots of the Bessel function of
order $n$ (Fig.~\ref{entACDC}b). Super Bloch oscillations are observed in the
entropy dynamics for the cases of a slight detuning from the resonant
condition (Fig.~\ref{entACDC}c).
The chief findings of this section are that entanglement entropy offers a useful alternative perspective for
each of the phenomena associated with the application of a general (static or dynamic) electric field.

\section{Summary and Conclusions}
To summarize, we studied the dynamics of many body entanglement in a
nearest neighbour tight binding chain with different forms of electric
field. For the static electric field we find that the dynamics
captures the well known Bloch oscillations with the constraint that
the system size must be greater than the Bloch length. A system size
scaling of the entanglement entropy at $t=nT_B$ shows a transition
from a DC regime to Bloch oscillation regime. For an AC field we find
that the entropy oscillates with the driving frequency at the
dynamically localized point whereas at other points it is
unbounded. For a combined AC+DC form of the field we find that at
resonantly tuned DC field, the unboundedness of entanglement entropy
verifies the coherent destruction of the localization caused by DC
field whereas a slight detuning from the resonance condition leads to
super Bloch oscillations in the entropy dynamics. From the system size
scaling of the super Bloch oscillations, we find a transition from DC
regime to super Bloch oscillations regime which demands a minimum
system size to observe super Bloch oscillations.
    
Our results in a closed system suggests that if one has to study
super Bloch oscillations in an open system which includes a set-up
where a device is coupled to the metallic leads, the device size has
to be greater than the super Bloch length to observe super Bloch
oscillations. We believe that the same experimental set-up, as put
forward in Ref.~\onlinecite{popescu2017emergence} can be generalized for this
purpose. Furthermore, time-dependent phenomena in a nonequilibrium set-up have mostly been
studied with a small number of degrees of freedom (like a quantum dot
for example). Therefore, it would be interesting to build on our current
work and extend it to study non-equilibrium properties of the nearest-neighbour
chain that is subjected to a time-dependent electric field.

\acknowledgments We thank Vikram Tripathi for useful discussions. A.S
is grateful to SERB for the startup grant (File Number:
YSS/2015/001696). D.S.B. acknowledges University Grants Commission (UGC), India for his Ph.D. fellowship.

\bibliography{ref} \onecolumngrid \appendix
\section{Time evolution of any general quantum State}
Let the initial thermal state at inverse temperature $\beta$ corresponding to Hamiltonian $H_{0}$ be
\begin{equation}
\rho\left(0\right)=\frac{e^{-\beta H_{0}}}{Z},
\end{equation}
where, $Z=\sum_{\alpha}e^{-\beta\epsilon_{\alpha}}$. This can be express in the eigenbasis $|\Psi_{n}\rangle$ of the Hamiltonian $H_0$ as 
\begin{equation}
\rho\left(0\right)= \frac{1}{Z}\sum_{n}e^{-\beta\epsilon_{n}}|\Psi_{n}\rangle\langle\Psi_{n}|.
\end{equation} 
At $t=0$, a parameter of the Hamiltonian is suddenly changed. The time evolution of the initial thermal state is now govern by new Hamiltonian $H_{1}$ as 
\begin{equation}
\rho\left(t\right)=\frac{1}{Z}\sum_{n}e^{-\beta\epsilon_{n}}|\Psi_{n}(t)\rangle\langle\Psi_{n}(t)|,
\end{equation}
where $|\Psi_{n}(t)\rangle$ can be written in terms of the eigenstates of the new Hamiltonian $H_{1}$ as
\begin{eqnarray}
H_{1}|\Phi_{n}\rangle=E_{n}|\Phi_{n}\rangle  \ \ \ \ \ \ \ \ \ \ \ \ \ \ \ \\\
|\Psi_{n}(t)\rangle= e^{-iH_{1}t/\hbar}|\Psi_{n}\rangle=\sum_{m}|\Phi_{m}\rangle e^{-iE_{m}t/\hbar}\langle\Phi_{m}|\Psi_{n}\rangle,
\end{eqnarray}
where $E_n$ are the eigenvalues of $H_1$. Now the time evolution of the density matrix can be written as
\begin{equation}
\rho\left(t\right)=\frac{1}{Z}\sum_{n}e^{-\beta\epsilon_{n}}\left(\sum_{mm^\prime}|\Phi_{m}\rangle e^{-i(E_{m}-E_{m^\prime})t/\hbar}\langle\Phi_{m^\prime}| \ \langle\Phi_{m}|\Psi_{n}\rangle\langle\Psi_{n}|\Phi_{m^\prime}\rangle\right)
\end{equation} 
In case of quenching from no electric field to some finite value of field $F$ in the free Fermionic Hamiltonian, the eigenvalues of the final Hamiltonian are given by $E_{m}=maF$, where $m=0,\pm 1,\pm 2, .... $.Hence the time evolution of density matrix can be written as 
\begin{equation}
\rho\left(t\right)=\frac{1}{Z}\sum_{n}e^{-\beta\epsilon_{n}}\left(\sum_{mm^\prime}|\Phi_{m}\rangle e^{-ia\omega_{B}(m-m^\prime)t/\hbar}\langle\Phi_{m^\prime}| \ \langle\Phi_{m}|\Psi_{n}\rangle\langle\Psi_{n}|\Phi_{m^\prime}\rangle\right)
\end{equation} 
where, $\omega_{B}=aF/\hbar$ is the Bloch frequency. After a Bloch period $t=t+T_{B}$, we have 
\begin{eqnarray}
\rho\left(t+T_{B}\right) =\frac{1}{Z}\sum_{n}e^{-\beta\epsilon_{n}}\left(\sum_{mm^\prime}|\Phi_{m}\rangle e^{-ia\omega_{B}(m-m^\prime)t/\hbar}e^{2\pi i(m^\prime-m})\langle\Phi_{m^\prime}|\langle\Phi_{m}|\Psi_{n}\rangle\langle\Psi_{n}|\Phi_{m^\prime}\rangle\right)
\end{eqnarray} 
Since, $\exp(2\pi i(m^\prime -m))=1$, we get
\begin{equation}
\rho\left(t+T_{B}\right) =\rho\left(t\right).
\end{equation}
So, the density matrix is also periodic in time with the time period of Bloch oscillations.
\section{Time evolution of a many body state and time-dependent correlation matrix}
The initial Hamiltonian of the system is
\begin{equation}
H_{0}=-\frac{1}{2}\sum_{n=-\infty}^{\infty}\left(c_{n}^{\dagger}c_{n+1}+c_{n+1}^{\dagger}c_{n}\right)
\end{equation}
Diagonalizing the Hamiltonian $H_{0}$, by new Fermionic operators,
$b^{\dagger}_{k}=\sum_{j}c^{\dagger}_{j}B_{jk}$, we can take the initial state to be the ground state of $H_0$ as
\begin{equation}
|\psi(0)\rangle = \prod_{\kappa\in K} b_{\kappa}^\dagger|0\rangle.
\end{equation}
After introducing the gradient term, the new Hamiltonian is given by
\begin{equation}
H=-\frac{1}{2}\sum_{n=-\infty}^{\infty}\left(c_{n}^{\dagger}c_{n+1}+c_{n+1}^{\dagger}c_{n}\right)+F\sum_{n=-\infty}^{\infty}\left(n-\frac{1}{2}\right)c_{n}^{\dagger}c_{n}
\end{equation}
by introducing new Fermionic operators,$d^{\dagger}_{k}=\sum_{j}c^{\dagger}_{j}D_{jk}$, we can diagonalize the Hamiltonian $H$,
\begin{equation}
H=\sum_{k}\epsilon_{k}d^{\dagger}_{k}d_{k},
\end{equation} 
where $\epsilon_{k}=kaF=\hbar k\omega_{B}, k= 0,\pm 1 .... $,\ and $\omega_{B}$ is the Bloch frequency.\\
The time evolution of the operator $d^{\dagger}_{k}$ is given by
\begin{equation}
\frac{d}{dt}d^{\dagger}_{k}=\frac{i}{\hbar}\left[H,d^{\dagger}_{k}\right]=\frac{i}{\hbar}\epsilon_{k}d^{\dagger}_{k},
\end{equation}
which gives,
\begin{equation}
d^{\dagger}_{k}(t)=d^{\dagger}_{k}(0) \exp(i\epsilon_{k}t/\hbar). 
\end{equation}
Now the time evolution of other Fermionic operators can be written as 
\begin{eqnarray}
c^{\dagger}_{j}(t)=\sum_{m,n} c^{\dagger}_{n}(0)D_{nm}\exp(i\epsilon_{m}t/\hbar)D^*_{mj}\\
b^{\dagger}_{j}(t)=\sum_{j} c^{\dagger}_{n}(0)\sum_{m,n} D_{jm}\exp(i\epsilon_{m}t/\hbar)D^*_{mn}B_{nk},
\end{eqnarray}
or,\begin{equation}
b^{\dagger}_{k}(t)= \sum_{j}B_{jk}(t)c^{\dagger}_{j},\quad
B_{jk}(t)=\sum_{m,n}D_{jm}\exp(i\epsilon_{m}t)D^*_{mn}B_{nk}(0).
\end{equation}
So the time evolution of the ground state can be written as
\begin{eqnarray}
|\psi_{0}(t)\rangle = \prod_{\kappa\in K} b_{\kappa}^\dagger (t)|0\rangle\\
|\psi(t)\rangle = \prod_{\kappa\in K}\left(\sum_{j} c_{\kappa}^\dagger (0)B_{jk}(t)\right)|0\rangle
\end{eqnarray}
If we translate the time by a Bloch period $T_{B}=2\pi/\omega_{B}$, the state becomes
\begin{eqnarray}
|\psi(t+T_{B})\rangle = \prod_{\kappa\in K}\left(\sum_{j} c_{\kappa}^\dagger (0)B_{jk}(t+T_{B})\right)|0\rangle,
\end{eqnarray}
now,
\begin{eqnarray}
B_{jk}(t+T_{B})=\sum_{m,n}D_{jm}\exp(i\epsilon_{m}(t+T_{B})/\hbar)D^*_{mn}B_{nk}(0)=\sum_{m,n}D_{jm}\exp(i\epsilon_{m}t/\hbar)D^*_{mn}B_{nk}(0)\exp(2\pi im).
\end{eqnarray}
Since, $\exp(2\pi im)=1$, we have
\begin{equation}
B_{jk}(t+T_{B})= B_{jk}(t)
\end{equation}
Hence, the state $|\psi_{0}(t)\rangle$ is periodic in time with the time period equal to the Bloch period $T_{B}$.

The time dependent correlation matrix can be written as 
\begin{eqnarray}
C_{mn}(t)=\langle \psi(t)|c^{\dagger}_{m}c_{n}|\psi(t)\rangle=\langle \psi(0)|c^{\dagger}_{m}(t)c_{n}(t)|\psi(0)\rangle
\end{eqnarray}
using the time evolution of Fermionic operators $c^{\dagger}_{j}$ and $c_{j}$ we can write the expression for the time dependent correlation matrix as 
\begin{equation}
C_{mn}(t)=\sum_{pq}\sum_{p^{\prime}q^{\prime}} D_{qp}\exp(i\epsilon_{p}t)D^*_{pm} C_{qq^{\prime}}(0) D^*_{q^\prime p^\prime}\exp(-i\epsilon_{p^{\prime}}t)D_{p^\prime n}.
\end{equation}
Which can be simplifies to
\begin{equation}
C(t)=U^{\dagger}(t)C(0)U(t),
\end{equation} 
where $U_{jk}(t)=\sum_{n}D^*_{jn}\exp(-i\epsilon_{n}t)D_{nk}$ and matrix $D$ diagonalizes the final Hamiltonian.
\end{document}